\documentclass[superscriptaddress, reprint,nofootinbib, amsmath,amssymb, floatfix,]{revtex4-1}

\usepackage[  ]{graphicx} 
\usepackage{dcolumn} 
\usepackage{bm,subfigure} 
 
\newcommand{\rmi}{\mathrm{i}}

\newcommand{\point}{\raise0.7ex\hbox{.}}

\newcommand{\rmB}{\mathrm{B}}
  
\newcommand{\xr}{\mathrm{x}}
\newcommand{\yr}{\mathrm{y}}
\newcommand{\zr}{\mathrm{z}}

\begin{document}

\preprint{APS/123-QED}

\title{Metamaterial control of  stimulated Brillouin scattering}

\author{M. J. A. Smith}
\email{m.smith@physics.usyd.edu.au}
\affiliation{Centre for Ultrahigh bandwidth Devices for Optical Systems (CUDOS) and Institute of Photonics and Optical Science (IPOS), School of Physics, The University of Sydney, NSW 2006, Australia}
 \affiliation{Centre for Ultrahigh bandwidth Devices for Optical Systems (CUDOS), School of Mathematical and Physical Sciences, University of Technology Sydney, NSW 2007, Australia }
\author{B. T. Kuhlmey}
\affiliation{Centre for Ultrahigh bandwidth Devices for Optical Systems (CUDOS) and Institute of Photonics and Optical Science (IPOS), School of Physics, The University of Sydney, NSW 2006, Australia}
\author{C. Martijn de Sterke}
\affiliation{Centre for Ultrahigh bandwidth Devices for Optical Systems (CUDOS) and Institute of Photonics and Optical Science (IPOS), School of Physics, The University of Sydney, NSW 2006, Australia}

\author{C. Wolff}
\author{M. Lapine}
\author{C. G. Poulton}
 \affiliation{Centre for Ultrahigh bandwidth Devices for Optical Systems (CUDOS), School of Mathematical and Physical Sciences, University of Technology Sydney, NSW 2007, Australia }

\date{\today} 

\begin{abstract} \noindent
Using full opto-acoustic  numerical simulations,  we demonstrate enhancement and suppression of the SBS gain   in a metamaterial comprising  a subwavelength cubic array of dielectric spheres  suspended in a dielectric background material. We develop a general theoretical framework and present several numerical examples using technologically important materials. For  As$_2$S$_3$ spheres   in  silicon, we achieve  a gain enhancement of    more than an order of magnitude   compared to pure silicon, and for GaAs spheres in silicon,  full suppression   is   obtained. The gain   for As$_2$S$_3$  glass    can also be strongly suppressed by embedding  silica spheres. The constituent terms of the gain coefficient  are shown to depend in a complex way on the filling fraction. We find that electrostriction is  the dominant effect behind the  control of SBS in bulk media.

\end{abstract}

\maketitle
  
   Stimulated Brillouin scattering (SBS)  is a nonlinear scattering process whereby an  incident electromagnetic pump field coherently drives an acoustic wave through the material, scattering the pump field and inducing a frequency shift in the returning, or Stokes, field \cite{boyd1990noise,eggleton2013inducing,powers2011fundamentals,kobyakov2010stimulated}.  This scattering process  features prominently in nonlinear optics, due to its relevance in the  design of nanoscale devices such as  on-chip tuneable photonic filters, Brillouin lasers and sensors \cite{eggleton2013inducing}.  That said, SBS is also regarded as a nuisance in optical communications systems, with considerable effort being focused on   techniques for its suppression \cite{peral1999degradation}.   The usual   process by which    sound  is coherently excited in the medium for SBS is   electrostriction      \cite{auld1973acoustic,wolff2014stimulated}, which describes when  an electric field induces       a      strain field in the material. These strains can  coherently drive   a    longitudinal  acoustic wave through the medium, inducing a periodic variation in the optical properties of the material (via the photoelastic effect). It is the combination of electrostriction and photoelasticity  which  scatters the incident optical field  \cite{eggleton2013inducing,auld1973acoustic,peral1999degradation};  thus, materials with strong photoelastic and electrostrictive properties   also tend to exhibit strong SBS. Conversely, materials with weak electrostriction and photoelasticity are poor candidates for experimental demonstrations of SBS.

A recent theoretical study by the  authors \cite{smith2015electrostriction}    demonstrated that the electrostrictive response of a material can be considerably enhanced   or suppressed through the introduction of    a subwavelength cubic array of spheres in a background material. Experimental work on  doped silica fibres has also shown considerable promise \cite{dragic2012sapphire}. Such     results   suggest that  SBS  is  also affected    through        subwavelength structuring. This is the motivation behind our study  into how metamaterial structuring   influences the SBS response.    In our work, we define an optoacoustic metamaterial as a structured material with period much smaller than both the acoustic and optical wavelengths in the material.  

In this framework, we rigorously model the electrostrictive response of a metamaterial using   perturbation procedures, with very few restrictive assumptions. The metamaterial we consider is a  cubic array of spheres embedded in a background material. In contrast to existing work  \cite{smith2015electrostriction} which gave an   analytical expression for the  electrostriction within a hydrostatic approximation, we incorporate shear effects in our model (which are generally non-negligible in   solid media but can easily be omitted for liquids). To evaluate the  SBS gain, we also evaluate other photonic and acoustic parameters in the subwavelength limit,  and  incorporate the effects of acoustic loss. To the best of our knowledge, this  is the  first investigation on the SBS behaviour of metamaterials and we emphasise that the results presented are for  intrinsic, or  bulk, SBS properties.  We   theoretically obtain       values for all  parameters  which feature    in the SBS gain coefficient, all of which vary differently as we tune the filling fraction of our metamaterial and    investigate   different material combinations.  

\begin{table*}[t]
 \centering
 \caption{\label{tab:table1}  \bf 
Bulk  parameters   
at $\lambda_1 = 1550 \, \mathrm{nm}$:    refractive index $n$,  photoelastic tensor coefficients  $p_{ij}$,      Brillouin frequency shift $\Omega_\mathrm{B}/(2 \pi)$ (units of  $\left[\mathrm{GHz}\right]$),  Brillouin   linewidth $\Gamma_\mathrm{B} / (2\pi)$ (units of $\left[\mathrm{MHz}\right]$),   gain coefficient $g_\mathrm{P}$ (units of $ \left[\mathrm{m} \cdot \mathrm{W}^{-1}\right]$),    phonon viscosity coefficients $\eta_{ij}$  (units of $\left[ \mathrm{mPa} \cdot \mathrm{s} \right]$),  stiffness tensor coefficients $C_{ij}$ (units of $\left[ \mathrm{GPa} \right]$),   material density $\rho$ (units of $\left[ \mathrm{kg} \cdot \mathrm{m}^{-3}\right]$), and     acoustic velocity $V_A$ (units of $\left[ \mathrm{m} \cdot \mathrm{s}^{-1} \right]$) \cite{weber2002handbook,abedin2005observation,rouvaen1974acoustic,galkiewicz1972photoelastic,pant2011chip,biegelsen1974photoelastic,helme1978phonon}, where $\dagger$ are  theoretical   estimates, and subscripts are in Voigt form.}
 
 { \small
 \begin{tabular}{lc |  ccc      | ccl |ccc| ccc|cc } \hline
Material   &$n$&  $p_{11}$ & $p_{12}$ &  $p_{44}$ & $\dfrac{\Omega_\mathrm{B}}{2\pi}$  & $\dfrac{\Gamma_\mathrm{B}}{2\pi}$  & $\mathrm{max} \, (g_\mathrm{P})$ & $\eta_{11}$ & $\eta_{12}$ & $\eta_{44}$ & $C_{11}$   & $C_{12}$ & $C_{44}$  & $\rho$   & $V_\mathrm{A}$  \\
\hline
  Fused SiO$_2$             &1.45  &0.12&  0.27 	& -0.075    	& 11.1&16& $4.52 \times 10^{-11}$   & 1.6$^\dagger$   & 1.29$^\dagger$  & 0.16$^\dagger$  &78.6 & 16.1 & 31.2 &2200 &5960     \\ 
 As$_2$S$_3$   &2.37 &0.25& 0.24 & 0.005     	& 7.95& 34&   $7.4 \times 10^{-10}$  & 1.8$^\dagger$   & 1.45$^\dagger$   & 0.18$^\dagger$   & 18.7 & 6.1 & 6.4  &3200 &2595      \\
 Si 	$\left[100\right]$  	  &3.48 &-0.094 & \!\! 0.017 & -0.051 &  38$^\dagger$ & 320$^\dagger$ & $ 2.4 \times 10^{-12} {}$$^\dagger$& 5.9   & 5.16   & 0.62   & 165.6 & 63.9 & 79.5 &2329 &8433     \\
GaAs $\left[100\right]$ &3.37 & -0.165& -0.14& -0.072 &   21$^\dagger$ & 167$^\dagger$  & $2.0 \times 10^{-10}$$^\dagger$ & 7.49   & 6.57   & 0.72   & 119 & 53.4 & 59.6 & 5320 & 4734   \\  
 \hline
\end{tabular}
}

 \end{table*}
  In addition to outlining  a general theoretical framework, we present numerical results for a selection of  silicon and chalcogenide glass-based metamaterials, to demonstrate suppression and enhancement of the intrinsic SBS gain coefficient. There is considerable interest in  silicon-based   materials due to a wide range of potential   applications in the electronics industry, as silicon is     CMOS-compatible      \cite{eggleton2013inducing}. That said, the biggest drawbacks in the use of conventional silicon as an SBS material are its inherently poor SBS gain coefficient, high speed of sound, and its large acoustic losses. We overcome these issues by introducing a suspension of   spheres in the background material, and demonstrate an order of magnitude enhancement in the  gain coefficient of bulk silicon using chalcogenide glass spheres. We also show  absolute suppression of     SBS in silicon using GaAs  spheres. 

The   example we present for an As$_2$S$_3$ background material shows  strong suppression in  the SBS gain coefficient when structured with a cubic lattice of silica spheres, which means that SBS would be observed at  much higher laser powers (i.e. this increases the SBS threshold). Demonstrating SBS suppression  in isotropic materials is relevant to those studying other nonlinear optical effects in   common laser glasses, such as four-wave mixing, where undesired SBS effects can dominate. 

 The procedure for deriving  the   coupled  intensity equations for      electrostriction-induced SBS  is well-known \cite{powers2011fundamentals,kobyakov2010stimulated,agrawal2007nonlinear}, and considers optical plane wave propagation in an isotropic bulk material. It   gives rise to the  SBS power gain spectrum      
\begin{equation}
\label{eq:gain}
g_\mathrm{P}= \frac{ 4 \pi^2 \gamma_\mathrm{  }^2   }{n  c \lambda_1^2 \rho   V_\mathrm{A} \Gamma^\mathrm{ }_\rmB} \left( \frac{( \Gamma^\mathrm{ }_\rmB/2)^2}{(\Omega_\rmB^\mathrm{ } - \Omega)^2 + ( \Gamma^\mathrm{ }_\rmB/2)^2}\right),
\end{equation}
 where  $\gamma $ is a measure  of the electrostrictive stress  in the medium (defined precisely below),  $n$ is the refractive index,  $c$ is  the speed of light in vacuum,   $\lambda_1$  is the incident  optical wavelength in vacuum, $\rho$ is  the mean material density,  $V_\mathrm{A}$  is the longitudinal acoustic wave velocity,  $\Omega$ is the angular frequency of the acoustic wave, $\Omega_\rmB /(2\pi)$ denotes the Brillouin frequency shift, and $\Gamma_\rmB$  is the Brillouin line width at half maximum,  with respect to angular frequency.   Note that  a conventional backwards SBS process    has $\Omega_\rmB =   q_\rmB  V_\mathrm{A}   \approx 2 \omega_1 n  V_\mathrm{A} /c$     where   $\mathbf{q}_\rmB = 2 \mathbf{k}_1$ is the corresponding wave vector \cite{powers2011fundamentals,kobyakov2010stimulated} and $\omega_1$ is the angular frequency of the incident optical  field.  The expression for $\gamma$   is   given in terms of the photoelastic tensor $p_{ijkl}$  \cite{newnham2004properties}, which  is defined in Einstein notation by
\begin{align} \label{eq:1}
\Delta \! \left( \varepsilon_{ij}^{-1} \right) &= p_{ijkl} s_{kl},
\end{align}
where  $\varepsilon_{ij}$ is the relative permittivity tensor,  $s_{ij} = \frac{1}{2} \left( \partial_i u_j + \partial_j u_i \right)$ is the strain tensor, $u_i$ is the elastic displacement from equilibrium, and $\Delta$ denotes the change resulting from the strain.   In this setting    we have \cite{peral1999degradation,rakich2010tailoring}
\begin{equation}
\label{eq:gamma12}
\gamma = \gamma_{\xr \xr \yr \yr} =  \varepsilon_\mathrm{r}^2 p_{\xr\xr\yr\yr}.
\end{equation} 
Consequently, provided we obtain values for all terms in \eqref{eq:gain}, then   the gain coefficient can be determined and  the SBS properties   of a metamaterial are characterised.  For reference, a range of   material parameters  \cite{weber2002handbook,abedin2005observation,rouvaen1974acoustic,galkiewicz1972photoelastic,pant2011chip,biegelsen1974photoelastic,helme1978phonon} are shown in Table \ref{tab:table1}.

\begin{figure*}[t]
\centering
\includegraphics[width=0.383\linewidth]{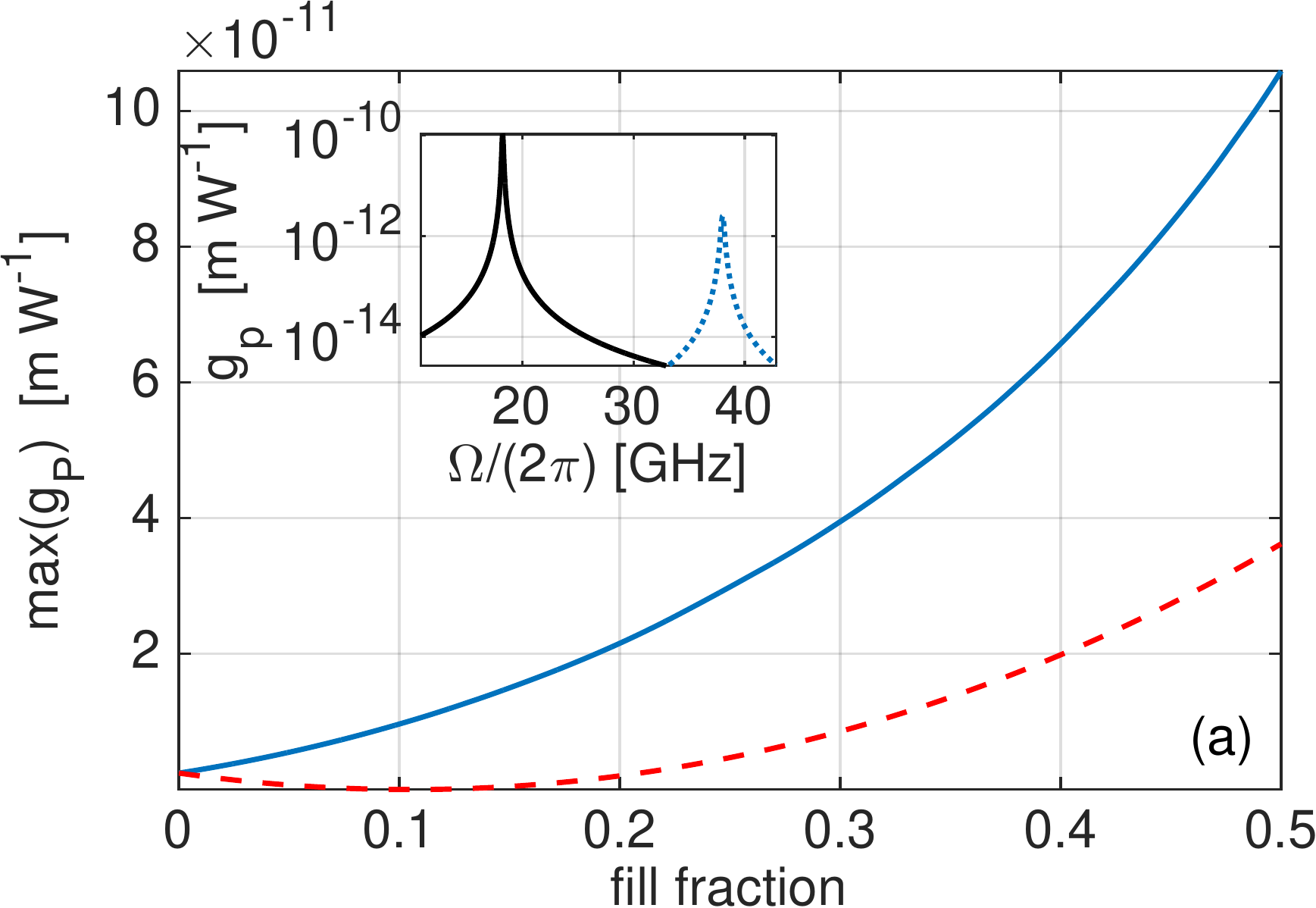} 
 \includegraphics[width=0.369\linewidth]{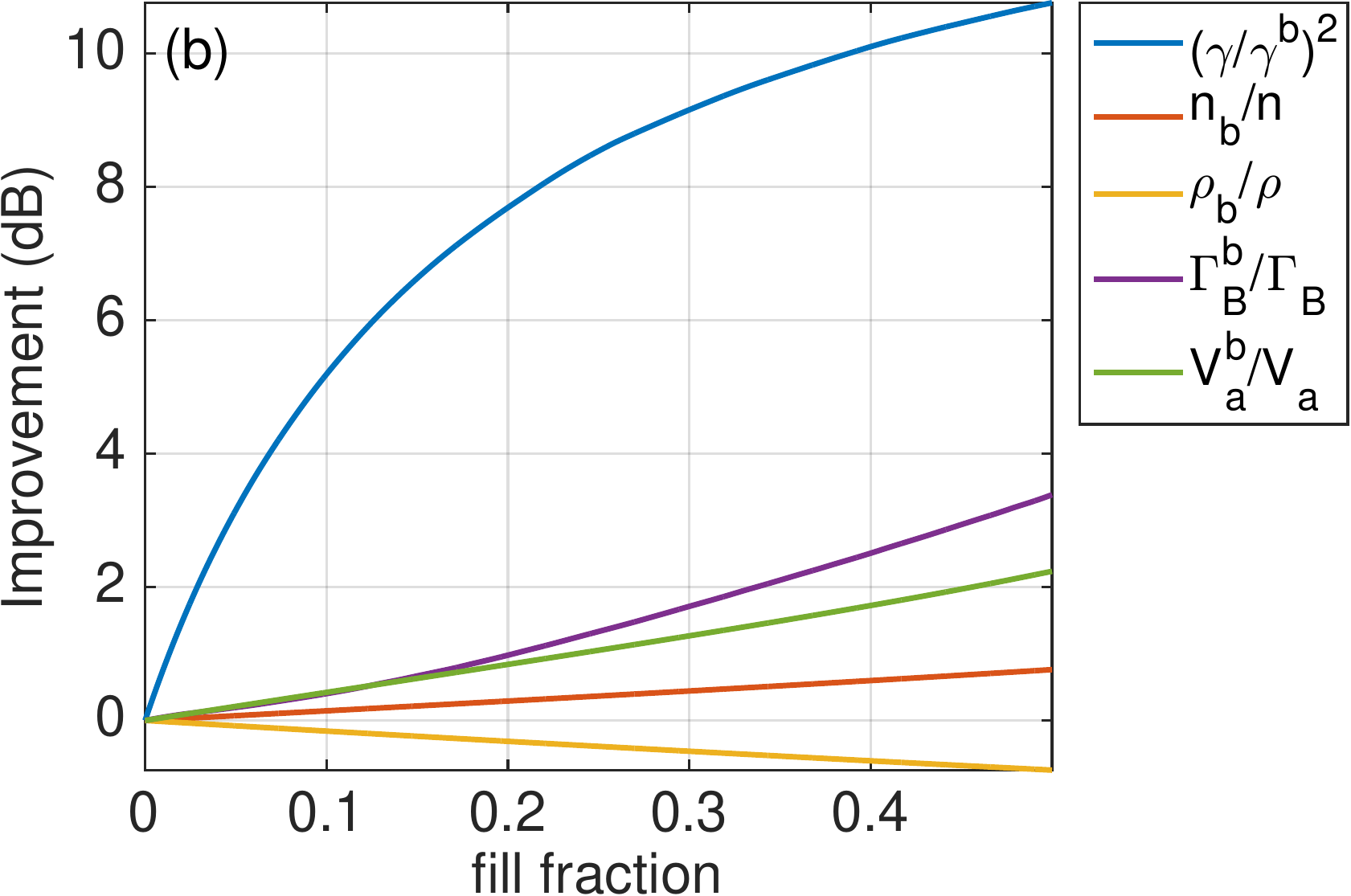}
\label{fig:2}
\caption{(a) Gain coefficient for   cubic lattice of As$_2$S$_3$ spheres in Si (blue) and GaAs spheres in Si (broken red) at $\lambda_1 = 1550$ nm for    $d = 50$ nm, inset:  gain coefficient  for pure Si (dotted blue) and cubic lattice of As$_2$S$_3$ spheres in Si at $f=50\%$ (black), (b) contribution from each term in \eqref{eq:gain} to   improvement  in $g_\mathrm{P}$     for As$_2$S$_3$ spheres in Si.}
\label{fig:1}
\end{figure*}

 We now proceed to obtain values  for all terms in \eqref{eq:gain}--\eqref{eq:gamma12},  beginning with an effective permittivity. Here, `effective' refers to a long-wavelength description of the properties of a metamaterial as if it were a uniform material.  
 
  For reference, we specify the unit cell to be symmetric about the origin, defining $d$ as the period of the cubic lattice, and $a$ as the radius of the sphere, from which we define the filling fraction as $f = 4 \pi a^3/ (3d^3)$. We remark that at dilute filling fractions the  choice of lattice geometry   is largely unimportant, although effects may be  pronounced at higher filling fractions. The effective permittivity tensor is obtained here   using a modification of the procedure outlined in \cite{bergman1978dielectric}, which is chosen for its conceptual simplicity and ease of numerical implementation (in this work, all problems are solved using a commercial finite element solver). This method involves first    solving the eigenvalue problem for Maxwell's equations  for a number of    Bloch vectors near the $\Gamma$ point. For each   vector we compute   the volume averaged   energy density \cite{jackson1962classical}
\begin{equation}
\label{eq:Uavg}
  U_\mathrm{avg} = \frac{1}{2} \frac{1}{V_\mathrm{WSC}} \varepsilon_0\langle \varepsilon_{ij}  {E}_i {E}_j^\ast \rangle ,
\end{equation}
where $E_j$ is the electric field distribution of the Bloch mode, $V_\mathrm{WSC}$ is the volume of the Wigner-Seitz cell, and $\langle \rangle$ denotes volume integration over the cell. This quantity is then equated to the   effective  energy density expression     
\begin{equation}
\label{eq:Ueff}
   U_\mathrm{eff}   =  \frac{1}{2} \frac{1}{\left(V_\mathrm{WSC}\right)^2}  \varepsilon_0  \varepsilon_{ij}^\mathrm{eff}     \langle  {E}_i\rangle  \langle  {E}_j\rangle^\ast   ,
   \end{equation}
   giving rise to a linear system that is solved directly for the effective permittivity tensor.  Following \eqref{eq:gamma12} we now determine the effective photoelastic constant $p_{\xr\xr\yr\yr}^\mathrm{eff}$. This is obtained by   mechanically perturbing the unit cell to approximate  a strain induced by a longitudinal acoustic wave propagating through the metamaterial. Thus, we solve   the acoustic wave equation \cite{auld1973acoustic} with zero body forces
\begin{equation}
\label{eq:acwaveeq}
-\rho \partial_t^2 u_i +
 \partial_j \left( C_{ijkl} \partial_k \right) u_l = 0, \quad \mbox{for} \quad i=\xr,\yr,\zr,
\end{equation}
inside the   unit cell, assuming we are  in the vicinity of $\Gamma$ (i.e. we  impose a time dependence of $\mathrm{exp}(-\rmi \Omega t)$ where $\Omega$ is in the long wavelength limit) where $C_{ijkl}$ denotes the stiffness tensor. To model the compression of the unit cell by the acoustic wave, we impose   the  boundary conditions   
\begin{equation}
\label{eq:strainbc}
u_j  \big|_{\partial W_{\pm \zr }}  = -D \zr \delta_{\zr j} \big|_{\partial W_{\pm \zr}}, \quad
u_j n_j\big|_{\partial W \backslash\left\{ \partial W_{\pm \zr} \right\}}  = 0  , 
\end{equation}
where $\partial W$ denotes the boundary of the entire unit cell, $D$ is the magnitude of the displacement, $n_j$ are the components of the local normal vector to the surface,  $\partial W_{\pm \zr}$ denote the faces of the cube   with normal vectors $n_j = \pm \delta_{\zr j}$, and $\delta_{ij}$ is the Kronecker delta. This boundary condition generates a compressed unit cell geometry and an internal strain field which modifies the constituent permittivity tensors,   making them spatially dependent (see \eqref{eq:1}).  Next, we repeat the procedure outlined in \eqref{eq:Uavg}--\eqref{eq:Ueff} to   obtain an  effective permittivity for the strained configuration, using the strained constituent permittivities. Having determined the  strained and unstrained  effective permittivity tensors (corresponding to  an imposed strain  over the   unit cell of $s_{\zr\zr} = -D$),     the   $p_{\xr\xr\yr\yr}^\mathrm{eff} $ coefficient  for the metamaterial follows directly  from the analogue to  \eqref{eq:1}, after using the symmetry properties of cubic crystals  \cite{nye1985physical} .

To determine the remaining terms in \eqref{eq:gain} we examine the  acoustic  properties of the unstrained metamaterial and  consider the acoustic wave equation \eqref{eq:acwaveeq} under the assumption of   time-harmonic fields taken in the long wavelength limit. In this setting, the  effective acoustic wave velocity is obtained   by solving the acoustic eigenvalue problem and evaluating $V_\mathrm{A}^\mathrm{eff}  = \tilde{\Omega} / \tilde{ {q}}$, where   $\tilde{\mathbf{q}} = 2 \mathbf{k}_1  = \left( 0,0, {4 \pi n_\mathrm{eff}}/{\lambda_1}\right)$ is the SBS resonant wave vector with corresponding acoustic frequency   $\tilde{\Omega}$ and       longitudinal  mode $\tilde{u}_j$. We calculate the effects of acoustic loss using perturbation theory; we substitute   $C_{ijkl}  + \eta_{ijkl} \partial_t$ for $C_{ijkl}$ in   \eqref{eq:acwaveeq}, where $\eta_{ijkl}$ is the phonon dynamic viscosity tensor \cite{auld1973acoustic,helme1978phonon}. Subsequently acoustic  frequencies are  perturbed as
\begin{equation}
\label{eq:Omega2ex}
\tilde{\Omega}^2 \rightarrow \tilde{\Omega}^2 + \rmi \tilde{\Omega} \frac{\langle  {a}_j  \tilde{  {u}}_j^\ast \rangle}{\langle \rho  \tilde{ {u}}_j   \tilde{  {u}}_j^\ast \rangle},
\end{equation}
 where 
$ a_i =\partial_j (\eta_{ijkl}  \partial_k \tilde{u}_l).$
 Numerically evaluating the square root of \eqref{eq:Omega2ex}   one obtains $\tilde{\Omega} \rightarrow \tilde{\Omega}_\mathrm{R} - \rmi \tilde{\Omega}_\mathrm{I}$ from which $\Omega_\mathrm{B} = \tilde{\Omega}_\mathrm{R}$ and $ \Gamma_\mathrm{B} = 2 \tilde{\Omega}_\mathrm{I}$ immediately  follows  \cite{jackson1962classical}.  We note that in order to evaluate the   linewidth of a metamaterial one must   possess the $\eta_{ijkl}$ of the constituent materials, and these are generally  not well-tabulated.  For uniform materials where $\eta_{ijkl}$ are not available, estimates are obtained by using results from     SBS experiments \cite{abedin2005observation,pant2011chip} and imposing $u_j = \mathrm{exp}\left( \rmi q \zr - \rmi \Omega t \right) \delta_{\zr j}$
to obtain  
$\tilde{\Omega}^2 \rightarrow \tilde{\Omega}^2 -   {\rmi \tilde{\Omega} q^2 \eta_{\zr\zr\zr\zr}}/{\rho}.$
 Taking the square root of both sides and evaluating a Taylor series in $q$ ultimately gives 
 \begin{equation}
 \eta_{\zr\zr\zr\zr} \approx \frac{V_\mathrm{A}^2 \Gamma_\mathrm{B} \rho}{\Omega_\mathrm{B}^2}.
 \end{equation}
 Following experimental data on $\eta_{ijkl}$  \cite{helme1978phonon} we     estimate $\eta_{\yr\zr\yr\zr}$ as being one order of magnitude smaller than $\eta_{\zr\zr\zr\zr}$, and assuming   the material is   isotropic, the identity  $\eta_{\yr\zr\yr\zr}= \frac{1}{2} \left( \eta_{\xr\xr\xr\xr} - \eta_{\xr\xr\yr\yr} \right)$ gives $\eta_{\xr\xr\yr\yr}$.   Note that  estimated values presented in Table \ref{tab:table1} are denoted by $\dagger$.

\begin{figure*}[t]
\centering
 \includegraphics[width=0.2886\linewidth]{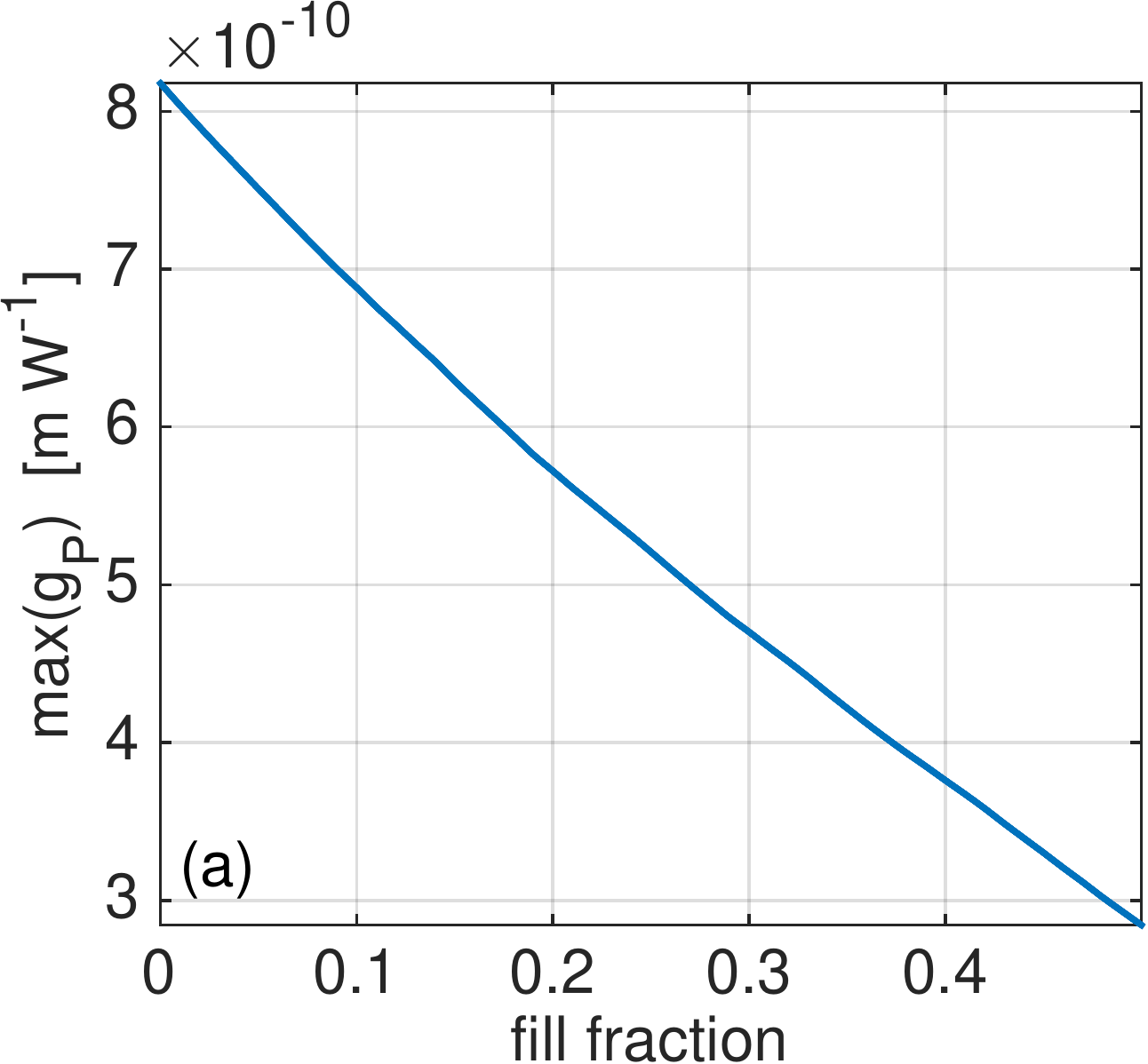}
\includegraphics[width=0.3693\linewidth]{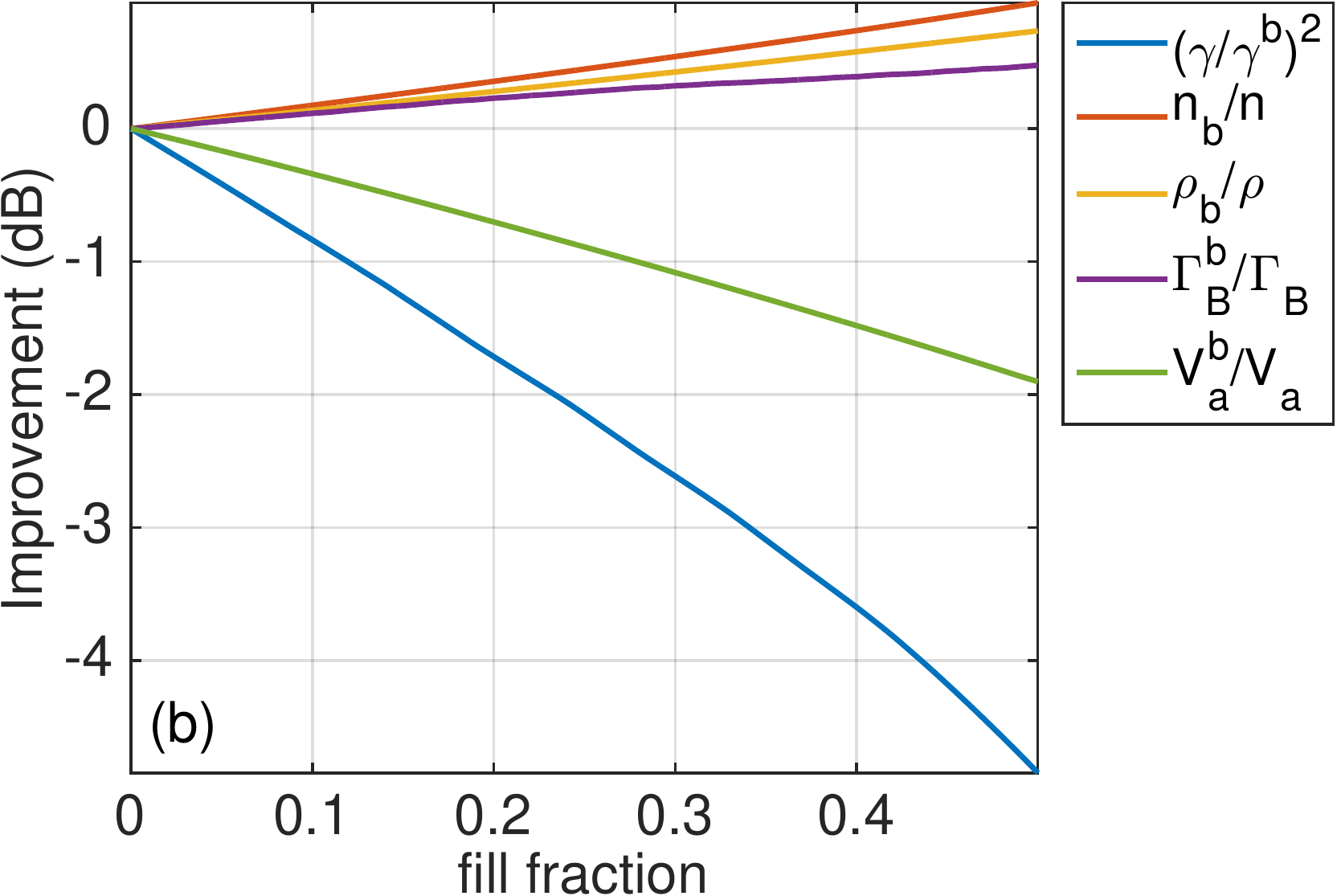}
 \includegraphics[width=0.2817\linewidth]{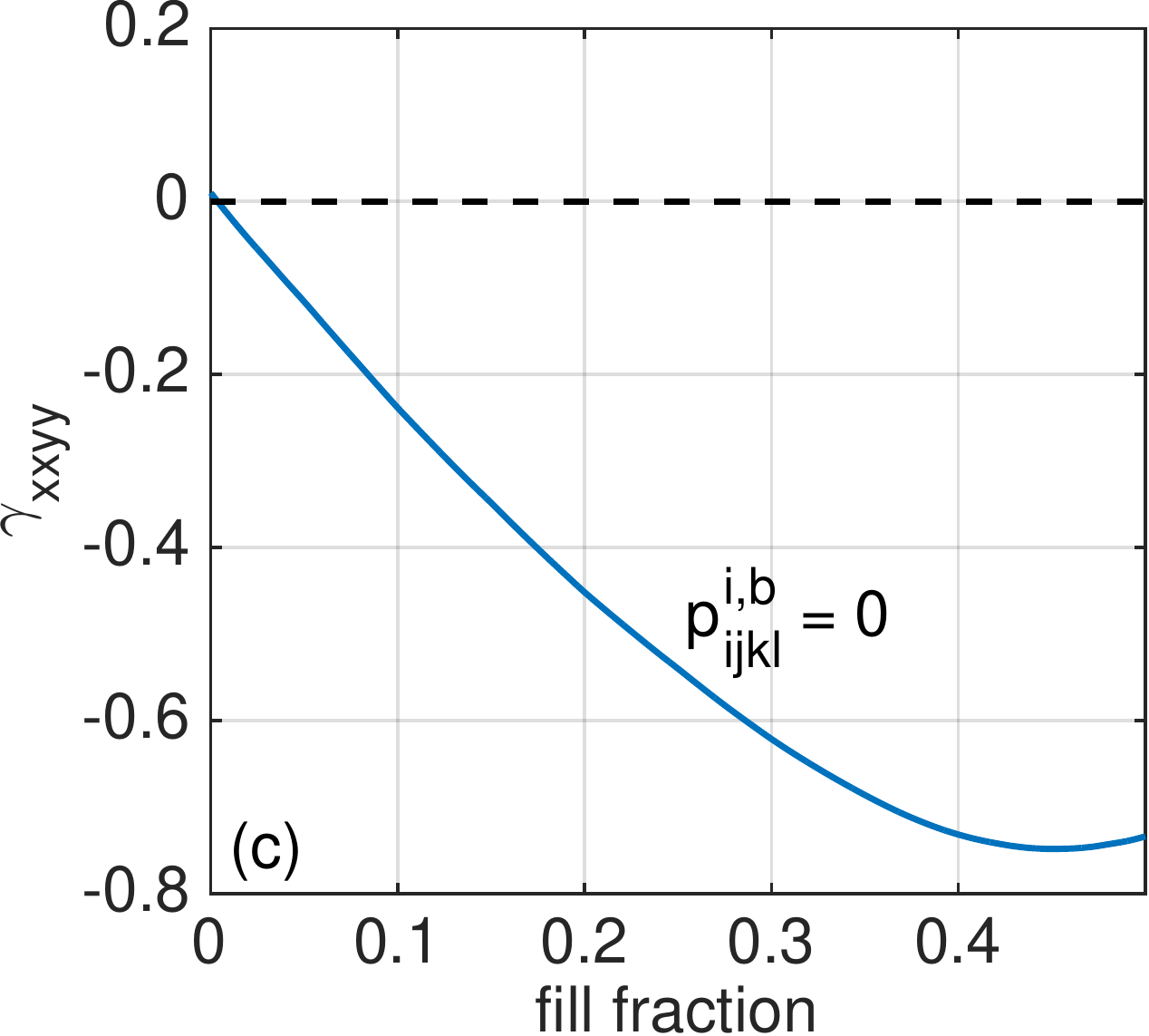}

\caption{(a) Gain coefficient for a cubic lattice of SiO$_2$ spheres in As$_2$S$_3$  at $\lambda_1 = 1550$ nm for $d = 50$ nm, (b)  contribution from each term in \eqref{eq:gain} to    improvement  in $g_\mathrm{P}$     for  SiO$_2$ spheres in As$_2$S$_3$, (c) electrostriction parameter $\gamma_\mathrm{xxyy}$ for SiO$_2$ spheres in As$_2$S$_3$   when the inclusion and background material photoelastic tensors   set to zero, showing a nonzero artificial contribution    \cite{smith2015electrostriction}.}
\label{fig:1a}
\end{figure*}

Having described the numerical procedures for determining all terms in \eqref{eq:gain} for the metamaterial, we now consider a selection of   illustrative examples. For each choice of pairwise material combination, we consider the maximum gain coefficient  \eqref{eq:gain}    against the filling fraction (where the maximum filling fraction for a cubic lattice of spheres is $f = \pi/6 \approx 0.52$). We also consider  how each parameter in   \eqref{eq:gain}    contributes to the   gain coefficient   by evaluating $$  10 \log_{10}\left(\frac{\mathrm{max}(g_\mathrm{P}) }{\mathrm{max}(g_\mathrm{P}^\mathrm{b} )}\right) =  
 10 \log_{10}\left( \left(\frac{\gamma}{\gamma^\mathrm{b}}\right)^2\right)  +  \ldots, $$
and superposing a plot of all    logarithmic terms in a single figure (where  $\mathrm{b}$ denotes the background material). In this way, the contribution  from each term is apparent because the improvement in the gain coefficient (in dB) is   then the sum of each curve value  at a given filling fraction. In Figure \ref{fig:1}, we present the   gain coefficient for a cubic lattice of As$_2$S$_3$ spheres in Si  at $\lambda_1 = 1550$ nm, where the  lattice period is $d = 50 \, \mathrm{nm}$  (solid blue curve). The period of the lattice is   chosen   to ensure that the structuring is both optically and acoustically subwavelength for all fill fractions: for the examples considered here, we have approximately 10 unit cells per optical wavelength.  From Figure \ref{fig:1}  we   see that As$_2$S$_3$ spheres in Si gives an order of magnitude enhancement in the SBS gain (\eqref{eq:gain})  from the bulk Si value shown in Table \ref{tab:table1}. In this case, an enhancement factor of more than 40 is achieved at  $f=50\%$ where $\mathrm{max}(g_\mathrm{P}) = 1.06 \times 10^{-10} \, \mathrm{m}\cdot\mathrm{W}^{-1}$, which is more than double that of pure fused SiO$_2$ (here we also have $\Gamma_\rmB / (2 \pi)= 147$ MHz and $\Omega_\rmB / (2 \pi)= 18$ GHz). The inset of Figure \ref{fig:1}a shows the gain spectrum for silicon at $f=0\%$ (dotted blue) and $f=50\%$ (black) for the above example, where the enhancement and shift is   visible.  In Figure \ref{fig:1}b, we observe that the SBS gain enhancement is   largely driven by an increase in   electrostriction, which is   greater than the contributions from improvements in the refractive index, acoustic velocity, and Brillouin linewidth combined. Note that the increasing density of the metamaterial drives a decrease in the gain coefficient, but for this example, only slightly mitigates the improvements arising from the other parameters. 

Also shown in Figure \ref{fig:1} is the   gain coefficient for GaAs spheres in Si where complete suppression of SBS is achieved at a filling fraction of $f = 10\%$, and   a $\mathrm{max}(g_\mathrm{P}) = 3.6 \times 10^{-11} \, \mathrm{m}\cdot\mathrm{W}^{-1}$ is achieved at $f = 50\%$. Note that at $f=50\%$ we have   structured Si with GaAs to   obtain a gain coefficient comparable to pure fused SiO$_2$, albeit with a broader linewidth of $\Gamma_\mathrm{B} / (2 \pi) = 223$ MHz and  a greater frequency shift of $\Omega_\rmB / (2 \pi)= 27$ GHz.  Analogously to the   example with  As$_2$S$_3$ spheres in Si, the SBS gain for GaAs spheres in Si for $f>10\%$ is driven by enhancements in all parameters except for the effective  material density (not shown). The $g_\mathrm{P} = 0$ observed at $f=10\%$ is  caused  by  $p_{\xr\xr\yr\yr}^\mathrm{eff} =0$, which is in turn  due to a sign change in   constituent $p_{\xr\xr\yr\yr}$ values (see Table \ref{tab:table1}).   In Figure \ref{fig:1a}a we show the   gain coefficient for SiO$_2$ spheres in As$_2$S$_3$, which demonstrates a more than 60\% suppression in the gain coefficient at $f=50\%$ (with corresponding values $\mathrm{max}(g_\mathrm{P}) =  2.8\times 10^{-10} \, \mathrm{m}\cdot\mathrm{W}^{-1}$, $\Gamma_\rmB / (2\pi) = 30$ MHz and $\Omega_\rmB / (2\pi) = 9$ GHz). The explanation for this suppression is found in Figure \ref{fig:1a}b where reductions in the   electrostriction and acoustic velocity outstrip positive contributions from all other remaining parameters. This points to  the acoustic velocity playing an   important role in the suppression of SBS in metamaterials, in addition to the electrostriction. Note our calculated value for the  SBS gain coefficient of As$_2$S$_3$ (i.e.,  at $f=0\%$) is within 10\% of the experimental value     in Table \ref{tab:table1}.

 In Figure \ref{fig:1a}c we show $\gamma_\mathrm{xxyy}$ for the same configuration but when the photoelastic tensors of the constituent materials are set to zero. In spite of this, the metamaterial has a non-vanishing electrostriction parameter. This ``artificial electrostriction'' \cite{smith2015electrostriction} arises from the different mechanical responses of the two constituent materials, and constitutes approximately $20 \%$ of the total $\gamma_\mathrm{xxyy}$   in Figure \ref{fig:1a}b at $f=50\%$. The presence of artificial electrostriction demonstrates that the properties of the metamaterial cannot be understood through direct mixing, even when the structuring is subwavelength.

In summary, we have shown that both  considerable enhancement and full suppression of SBS in silicon is achieved through a careful choice of inclusion material in a  metamaterial comprising spheres in a cubic lattice. Calculations (not discussed here) on face-centred cubic lattices of spheres indicate that the specific lattice geometry has a minimal effect on the gain in the dilute limit. SBS is a complicated process, involving optical and acoustic waves together with their mutual interaction. We have implemented a rigorous microscopic procedure which encompasses all contributing physical processes.

The enhancement of the silicon gain coefficient to   values greater than, or comparative to, fused silica is particularly promising for designers of small-scale, silicon based SBS devices. There is also considerable scope for metamaterials where the acoustic velocity contrast     and   the Brillouin linewidth contrast is high, as the contributions of these parameters have been shown here to play  an important role in controlling SBS. 
    
\section*{Acknowledgements}
This work was supported by the Australian Research Council (CUDOS Centre of Excellence, CE110001018).

\bibliography{SBS_meta_bib}

\end{document}